*Anatase steps trap electrons*

# Charge Trapping at the Step Edges of TiO$_2$ Anatase (101) **

*Martin Setvin, Xianfeng Hao, Benjamin Daniel, Jiri Pavelec, Zbynek Novotny, Gareth S. Parkinson, Michael Schmid, Georg Kresse, Cesare Franchini, and Ulrike Diebold\**

*A combination of Photoemission, Atomic Force and Scanning Tunneling Microscopy/Spectroscopy measurements shows that excess electrons in TiO$_2$ anatase (101) surface are trapped at step edges. Consequently, steps act as preferred adsorption sites for O$_2$. In Density Functional Theory calculations electrons localize at clean step edges, this tendency is enhanced by O vacancies and hydroxylation. The results show the importance of defects for the wide-ranging applications of titania.*

TiO$_2$ is used extensively in photocatalysis[1-2] and photoelectrochemical (Grätzel) solar cells,[3] and has promising properties for several other fields.[4-5] The material exists in two technologically important forms, rutile and anatase. While the fundamental surface properties of the stable rutile phase are well-understood,[6-7] less is known about the metastable anatase, which is used in most applications.

The behavior of excess electrons introduced into the material is a key issue in its applications. Electrons can be added by doping (oxygen vacancies (V$_O$s), Ti interstitials (Ti$_{int}$s), terminal hydroxyls, impurities),[6, 8] or UV light. These electrons can be trapped in the crystal lattice, forming a quasiparticle called the small polaron,[9] in essence a Ti$^{3+}$ ion with its surrounding lattice distortions. Transfer of electrons to adsorbed molecules is an elementary step in reduction and oxidation reactions. As a consequence, the charge-trapping centers, which accumulate excess electrons, often act as preferred sites for adsorption[1, 6, 10] and chemical reactions. electron trapping in a perfect anatase lattice is not favorable – neither in the bulk, nor on the (101) surface.[11] This is in stark contrast to the rutile (110) surface, which allows self-trapping of electrons at any surface or subsurface Ti atom.[12-14] In anatase, trapping can only occur at defects; this paper investigates the location of these trapped electrons.

Even though small polaron formation at the anatase (101) surface was previously excluded,[11] photoemission data always show a significant gap state ~1 eV below the Fermi level (E$_F$),[15] characteristic of a trapped electron.[11] Anatase(101) surfaces prepared under UHV conditions do not contain surface V$_O$s[16] or other visible point defects, thus the origin of this state deserves a deeper analysis. We note that the (001) plane of anatase does not show the 1-eV gap state in photoemission[15] and a delocalized, "large polaron"

[*] Martin Setvin, Xienfeng Hao, Jiri Pavelec, Zbynek Novotny, Gareth S. Parkinson, Michael Schmid, Ulrike Diebold
Institute of Applied Physics
Vienna University of Technology
Wiedner Hauptstrasse 8-10/134, 1040 Vienna, Austria
E-mail: diebold@iap.tuwien.ac.at

Cesare Franchini, Georg Kresse
Faculty of Physics and Center for Comp. Materials Science
Universität Wien, Sensengasse 8/12, A-1090 Wien, Austria

Martin Setvin
Institute of Physics, Academy of Sciences of the Czech Republic, Cukrovarnicka 10, 16200, Czech Republic

[**] The work was supported by the Austrian Science Fund (FWF; project F45) and the ERC grant "Oxide Surfaces". We thank HZB for the allocation of synchrotron radiation beamtime.

Supporting information for this article is available on the WWW under http://www.angewandte.org.

(manifested by a peak at ~40 meV below E$_F$ in photoemission) was reported there.[17] Here we use a combination of spatially resolved experimental techniques (STM, STS) and theoretical calculations (DFT) to resolve this issue. We show that the trapped electrons are located exclusively at the step edges of the anatase (101) surface. Importantly, the steps activate the surface for O$_2$ adsorption.

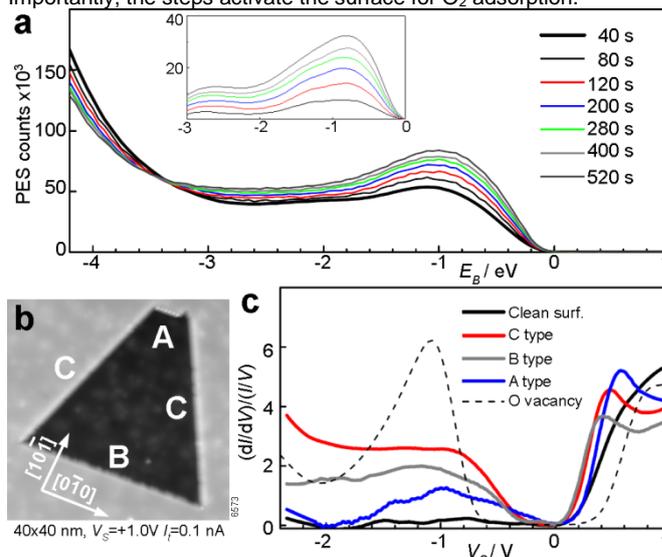

*Figure 1.* a) Photoemission spectra (hν = 130 eV, room temperature) of a TiO$_2$ anatase (101) surface taken after various exposures to the photon beam, measured at 300 K. The inset shows the same spectra after subtracting the first spectrum (40 s irradiation). b) Empty-states STM image (T = 78 K) of anatase (101) with a typical step edge configuration, labeled A-C according to ref. [18] c) STS spectra measured at the steps and at the flat surface.

Figure 1a shows photoemission spectra of the anatase (101) surface. No intensity is expected in the band gap region (0-3 eV below E$_F$). Instead a gap state is observed at a binding energy of (-1.0 ± 0.2) eV, similar to data reported earlier.[15] The pronounced tail behind this peak is unusual; it is not observed in photoemission of rutile (110). We will show that this tail does not stem from a background of inelastically scattered photoelectrons but is characteristic for electrons trapped at steps.

Even though as-prepared anatase(101) surfaces do not contain surface V$_O$s,[16, 19] these can be created by the photon beam and thus the possibility of beam damage needs to be taken into account. Figure 1a displays PE spectra after various irradiation times, ranging from 40 to 520 s. The inset in Figure 1a shows the difference between the first curve (40 s irradiation) and the curves after longer irradiation times. The evolution of the state in the difference spectra represents the signal coming from the beam-induced defects (likely surface V$_O$s). Note that the tail at higher binding energies is suppressed in these spectra. From this analysis we conclude that the initial gap state in Figure 1a (40 s) is too intense to be solely due to the beam damage; it must have a different origin.

To determine the spatial distribution of the gap state we investigated the surface with STM and STS, see Figures 1b,c. At negative bias voltages (filled states), the clean anatase surface has a very small local density of states (LDOS), as expected.[11] The situation is different when spectra are taken close to steps, however. Three different step types, termed A, B, and C as in a previous study,[18] appear on the anatase (101) surface; the difference in their formation energies results in terraces with a characteristic, trapezoidal shape (Figure 1b). STS spectra (Figure 1c) taken close to these steps show a significant LDOS within the band gap. The peaks have a maximum at ~1 eV below $E_F$ and a strong tail deeper in the band gap, very similar to the initial photoemission spectrum (Figure 1a, 40 s). STS shows the gap state directly at, as well as up to 2 lattice constants away from the step edge, at both, the upper and lower terrace. Spectra measured at various positions along the step edge are similar (see Figure S1 in the Supplement); the LDOS is homogeneous along the step. For comparison, we include an STS spectrum from an – artificially created [11, 19-20] – surface $V_O$ on anatase (101), see the dashed line in Figure 1c. We note a much more pronounced peak, similar to the ($V_O$-derived) difference spectra in the inset of Figure 1a. The LDOS is highest at the C-type step. Step B, which can be viewed as a small stripe of a (100) plane shows a much lower electron density.[18] Step A shows an intermediate intensity; this step is polar and possibly reconstructed.

To evaluate whether the charge trapping is an intrinsic property of the step (for example, whether a structurally perfect and clean step allows self-trapping of electrons excited by UV light) we employed DFT calculations. We took the C-type step as a model, as this step exhibits the strongest gap state in STS. When an excess electron was added to a (145)-oriented vicinal slab, electron localization at the step was slightly favored compared to a delocalized solution (polaronic energy[11-12, 23] ~46 meV, for details see the supplement). This value seems rather low but it is well in-line with polaron formation energies calculated for rutile.[11] One should further consider that the steps contain kinks and other defects (see Figure S3 in the Supplement), which locally increase the lattice flexibility and, in turn, further support polaron formation.

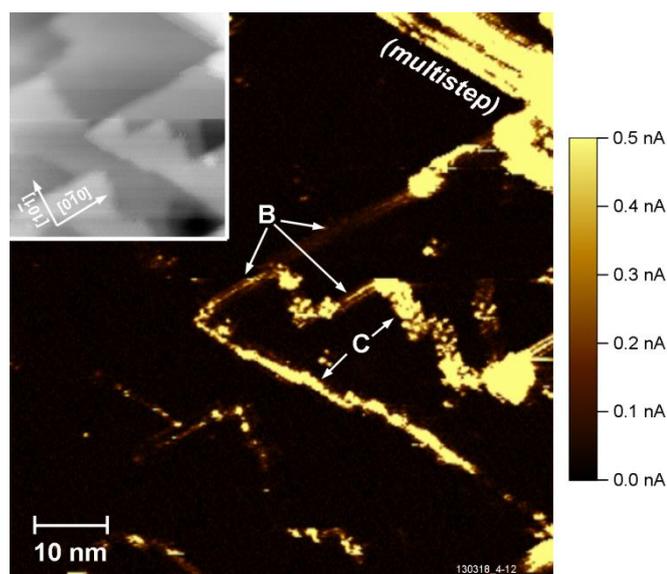

*Figure 2.* **Spatial distribution of filled states on anatase (101).** The inset shows a topographic nc–AFM image of the anatase (101) surface taken in constant frequency-shift mode ($V_{sample}$ = –1.5 V, $f_0$ = 59 kHz, A = 0.4 nm, Q = 1500, 300 K). The large image shows the tunneling current recorded simultaneously with the topography. It shows filled states, localized at step edges.

In constant height STM mode it is possible to image the filled DOS on flat anatase(101) terraces,[11] albeit not close to step edges where a feedback loop is needed for keeping the tip-sample distance. The tunneling-current-based feedback loop fails in filled states because the anatase (101) terraces have negligible LDOS in the band gap. Instead we resort to non-contact atomic force microscopy (nc-AFM), which allows the simultaneous measurement of force and tunneling current, see Figure 2. The measured force (more precisely, the frequency shift) is used to maintain the tip-sample distance. The corresponding topography image is shown in the inset of Figure 2. The sample was biased negatively (-1.5 V), thus the tunneling current shows the distribution of filled states. The current is below the detection limit at the terraces, and, apart from a few defects, the LDOS is strictly localized at the steps. The STS spectra in Figure 1c are relevant for quantifying the intensity of the filled LDOS, while Figure 2 shows its spatial distribution.

We use $O_2$ adsorption to illustrate how the step edges affect the chemical reactivity of anatase. $O_2$ is an electron scavenger and can chemisorb on $TiO_2$ only after accepting an extra electron from the surface.[1, 21] Figures 3 a, b show large-scale STM images of anatase (101) after $O_2$ exposure. The features at the terraces are due to different forms of adsorbed $O_2$ as described earlier.[20] The STM contrast at step edges undergoes a significant change even after small $O_2$ exposures. On the clean surface, in empty-states images the step edges always appear slightly brighter than the terraces, see Figure 1b. After dosing $O_2$, the steps become dark, which we attribute to upwards band bending caused by accumulation of negative charge.[22]

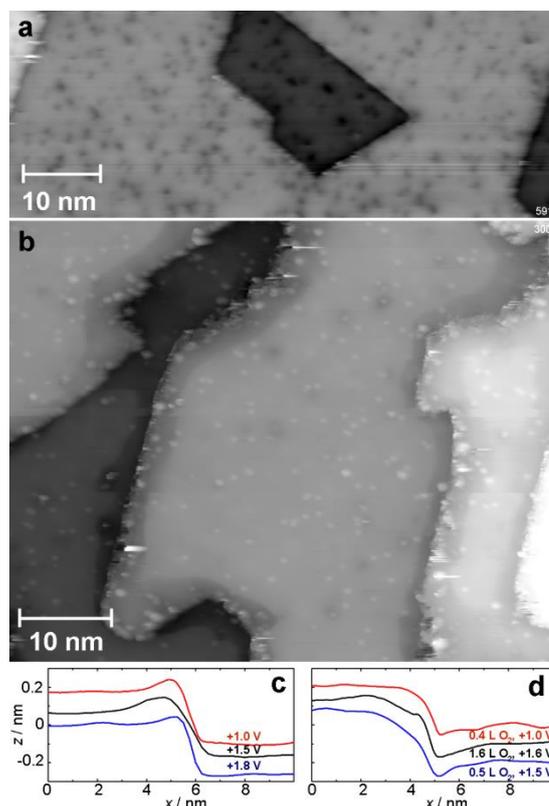

*Figure 3.* **Influence of the step–edge states on the adsorption of $O_2$** a) STM image ($V_s$ = +2.3 V, $I_t$ = 0.1 nA) of anatase (101) after dosing 0.4 Langmuir $O_2$ at T = 40 K. b) The surface after dosing 20 L $O_2$ at T=105 K, scanned at 6 K. ($V_s$ = +1.5 V, $I_t$ = 0.08 nA. In both cases, the significant upwards band-bending results in a dark contrast at the step edges and indicates adsorbed oxygen. c, d) Line profiles across C-type step edges of clean and $O_2$ - exposed surfaces, respectively. Each profile is from a different experimental session.

In our calculations, we also considered the presence of oxygen vacancies and hydroxyl groups at the steps. As detailed in the Supplement, both species are energetically possible at the steps and both provide electrons that trap at neighboring Ti sites. The presence of donors at the steps is further indicated by the experimental data: steps of a clean surface appear bright in empty-states STM images (see Figure 1b). For n-type semiconductors, this indicates the presence of donors[22]. For electrostatic reasons, it is likely that the electrons trapped at anatase steps originate mainly from the donors at the steps, although trapping of electrons from other sites, for example donors in the bulk, is also energetically possible as discussed above.

In summary, we have shown that, on the (101) surface of anatase, step edges are donors. This makes the steps preferential adsorption sites for acceptor molecules and, possibly, reactive centers for (photo-)catalytic reactions. Our results point to the great importance of step edges for $TiO_2$ anatase. Steps and line defects of catalytic materials such as rutile $TiO_2$, $CeO_2$ or MgO have been the focus of interest[24-27] and increased activities have been reported for certain reactions. The importance of the anatase steps is special due to the fact that charge trapping is not favorable on anatase (101) terraces and that the dominant (101) plane is relatively inert.

[1] M. A. Henderson, *Surf. Sci. Rep.* **2011**, *66*, 185-297.
[2] A. Linsebigler, G. Lu, J. R. Yates, *Chem. Rev.* **1995**, *95*, 735-798.
[3] M. Grätzel, *Nature* **2001**, *414*, 338-344.
[4] K. Szot, M. Rogala, W. Speier, Z. Klusek, A. Besmehn, R. Waser, *Nanotechnology* **2011**, *22*, 1-22.
[5] Y. Furubayashi, T. Hitosugi, Y. Yamamoto, K. Inaba, G. Kinoda, Y. Hirose, T. Shimada, T. Hasegewa, *Appl. Phys. Lett.* **2005**, *86*, 252101.
[6] U. Diebold, *Surf. Sci. Rep.* **2003**, *48*, 53-229.
[7] H. Cheng, A. Selloni, *The Journal of Chemical Physics* **2009**, *131*, 054703.
[8] S. Wendt, T. S. Sprunger, E. Lira, G. K. H. Madsen, Z. Li, J. Hansen, J. Matthiesen, A. Blekinge-Rasmussen, E. Laegsgaard, B. Hammer, F. Besenbacher, *Science* **2008**, *320*, 1755-1759.
[9] I. G. Austin, N. F. Mott, *Advanc. Phys.* **2001**, *50*, 758-812.
[10] Y. He, A. Tilocca, O. Dulub, A. Selloni, U. Diebold, *Nature Materials* **2009**, 585-589.
[11] M. Setvin, *Submitted* **2013**.
[12] A. Janotti, C. Franchini, J. B. Varley, G. Kresse, C. G. V. d. Valle, *Phys. Stat. Sol.* **2013**, *7*, 199-203.
[13] P. Krüger, S. Bourgeois, B. Domenichini, H. Magnan, D. Chandesris, P. L. Fevre, A. M. Flank, J. Jupille, L. Floreano, A. Cossaro, A. Verdini, A. Morgante, *Phys. Rev. Lett.* **2008**, *100*, 055501.
[14] P. M. Kowalski, M. F. Camellone, N. N. Nair, B. Meyer, D. Marx, *Phys. Rev. Lett.* **2010**, *105*, 146405.
[15] A. G. Thomas, W. R. Flavell, A. K. Mallick, A. R. Kumarasinghe, D. Tsoutsou, N. Khan, C. Chatwin, S. Rayner, G. C. Smith, R. L. Stockbauer, S. Warren, T. K. Johal, S. Patel, D. Holland, A. Taleb, F. Wiame, *Phys. Rev. B* **2007**, *75*, 035105.
[16] Y. He, O. Dulub, H. Cheng, A. Selloni, U. Diebold, *Phys. Rev. Lett.* **2009**, *102*, 106105.
[17] S. Moser, L. Moreschini, J. Jacimovic, O. S. Barsic, H. Berger, A. Magrez, Y. J. Chang, K. S. Kim, A. Bostwick, E. Rotenberg, L. Forro, M. Grioni, *Phys. Rev. Lett.* **2013**, *110*, 196403.
[18] X.-Q. Gong, A. Selloni, M. Batzill, U. Diebold, *Nature Materials* **2006**, *5*, 665-670.
[19] P. Scheiber, M. Fidler, O. Dulub, M. Schmid, U. Diebold, W. Hou, U. Aschauer, A. Selloni, *Phys. Rev. Lett.* **2012**, *109*, 136103.
[20] M. Setvin, U. Aschauer, P. Scheiber, Y.-F. Li, W. Hou, M. Schmid, A. Selloni, U. Diebold, *Science* **2013**, *341*, 988-991.
[21] Y. Cui, X. Shao, M. Baldofski, J. Sauer, N. Nilius, H.-J. Freund, *Angew. Chem. Int. Ed.* **2013**, *52*, 11385-11387, *Angew. Chem.* **2013**, *125*, 11595-11598.
[22] P. Ebert, *Surf. Sci. Rep.* **1999**, *33*, 121-303.
[23] I. G. Austin, N. F. Mott, *Adv. Phys.* **2001**, *50*, 757-812.
[24] U. Martinez, J. Hansen, E. Lira, H. H. Kristoffersen, P. Huo, R. Bechstein, E. Laegsgaard, F. Besenbacher, B. Hammer, S. Wendt, *Phys. Rev. Lett.* **2012**, *109*, 155501.
[25] H. Takahashi, R. Watanabe, Y. Miyauchi, G. Mizutani, *J. Chem. Phys.* **2011**, *134*, 154704.
[26] N. Nilius, S. M. Kozlov, J.-F. Jerratsch, M. Baron, X. Shao, F. Vines, S. Shaikhutdinov, K. M. Neyman, H.-J. Freund, *ACS Nano* **2012**, *6*, 1126-1133.
[27] G. Pacchioni, H. Freund, *Chem. Rev.* **2013**, *113*, 4035-4072.

**Entry for the Table of Contents**

*Anatase steps trap electrons*

Martin Setvin, Xianfeng Hao, Benjamin Daniel, Jiri Pavelec, Zbynek Novotny, Gareth Parkinson, Michael Schmid, Georg Kresse, Cesare Franchini, Ulrike Diebold

Charge trapping at the step edges of $TiO_2$ anatase (101)

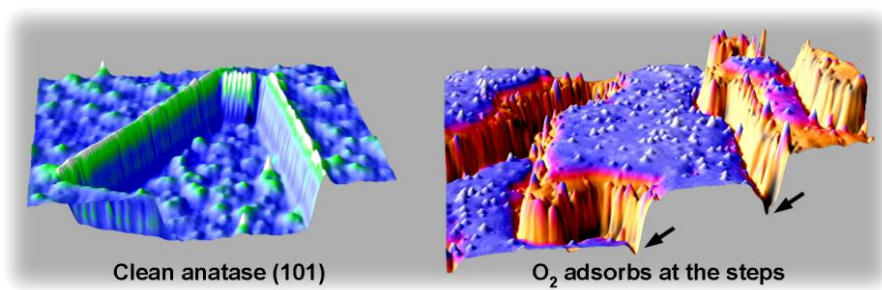

Step edges on $TiO_2$ anatase(101) surface act as exclusive charge trapping centers. While the electron trapping is not favorable at (101) terraces, it is possible at the steps. It results in higher reactivity of the steps towards some adsorbates. We illustrate this on an example of $O_2$ adsorption.